# Job Recommendation: Leveraging Progression of Job Applications

**Amber Nigam**
Georgia Institute of Technology, PeopleStrong
anigam9@gatech.edu

**Aakash Roy**
PeopleStrong
aakash.roy@peoplestrong.com

**Arpan Saxena**
TCS
arpansaxena17may@gmail.com

**Hartaran Singh**
PeopleStrong
hartaran.singh@peoplestrong.com

**Abstract**

Job recommendation has traditionally been treated as a filter-based match or as a recommendation based on the features of jobs and candidates as discrete entities. In this paper, we introduce a methodology where we leverage the progression of job selection by candidates using machine learning. Additionally, our recommendation is composed of several other sub-recommendations that contribute to at least one of a) making recommendations serendipitous for the end user b) overcoming cold-start for both candidates and jobs. One of the unique selling propositions of our methodology is the way we have used skills as embedded features and derived latent competencies from them, thereby attempting to expand the skills of candidates and jobs to achieve more coverage in the skill domain. We have deployed our model in a real-world job recommender system and have achieved the best click-through rate through a blended approach of machine-learned recommendations and other sub-recommendations. For recommending jobs through machine learning that forms a significant part of our recommendation, we achieve the best results through Bi-LSTM with attention.

## 1 Introduction

Job recommendation is primarily aimed at supporting the discovery of jobs that may interest the user. It should be dynamic in order to cater to the changing preferences of the user. For instance, a person looking for a job at *X* location might not be interested in the location a few years down the line. Similarly, a job that is relevant for an individual now might not be exciting enough in the future because of a possible upskilling. Consequently, this puts the onus of accounting for such variables on the recommender system to always be context-aware and relevant.

Jobs have criteria of suitability mentioned in the job description that a candidate is supposed to satisfy. Some jobs are more definite about the criteria than others, which might depend on attributes like *company* and *designation*. Certain features like *skills* and *designations* that have high dimensionality must be meticulously represented for algorithms to use them efficiently. Bastian ,et al., (2014) mentioned the importance of skills as an identifier of talent Like jobs, candidates also have a few attributes associated which hints us to the kind of jobs they might prefer. For instance, a candidate having proficiency in *HTML*, *JavaScript*, *Node.JS*, and *AWS* may prefer a *Full Stack Developer* job but may be lacking the explicit mention of the latter in the candidate's profile. Candidates also have a professional summary that includes features like *companies*, *job roles*, and *duration of work*. Candidate preferences are not static, and they may change as they progress in their career. Data describing the progression of candidates through their academic and professional careers might provide hints of their next steps, and thus it can be a good indicator of their motivations and preferences.

The match between candidates and jobs involves a complex amalgamation of the attributes of candidates and jobs. The intent is to identify the patterns in the data for recommending relevant jobs to candidates. The prediction of jobs for a candidate is based on information derived from the data about candidates applying for jobs on a web portal. A rule-based recommender system might not be an ideal solution as it is bound to miss cases, especially the nuances that humans cannot comprehend.

## 2 Related Work

Recommender systems have been extensively applied to suggest concise items of interest to the users and drive higher click-through rates (CTR) (Covington, et al., 2016 ; Gomez-Uribe, et al., 2016; Okura, et al., 2017). Video sharing website YouTube and media-services provider Netflix extensively use recommender systems to suggest videos and movies to their users respectively. 60% of videos watched on YouTube, and 80% of movies watched on Netflix are due to recommendations (Covington et al., 2016; Gomez-Uribe et al., 2016). In some approaches improvements in CTR using recommender systems are also favored (Elsafty, et al., 2018). Since early literature, recommender systems have been broadly categorized into content-based, collaborative filtering and hybrid, based on the features utilized in model input (Adomavicius and Tuzhilin, 2005). Recommender systems are often required to solve the cold-start problem where there may be insufficient information about the user, item or their interactions (Abel, et al., 2017; Chen,et al., 2019; Schein, et al., 2002). Recommender systems have also been applied in the field of recommending jobs to prospective employees (Chen, et al., 2019; Jiang,et al., 2019; Abel,et al., 2016). Elsafty et al.,(2018) used a document-based recommender system with dense representations and showed 8% relative increase. They used Word2Vec (Mikolov, et al., 2013) and Doc2Vec (Le and Mikolov, et al., 2014) to extract semantic relationships between jobs using job title and job descriptions.

Kenthapadi et al.,(2017) in their paper discussed the personalized job recommendation strategy at LinkedIn. They observed that the job recommendation problem has fundamental differences with other recommender systems involving books, movies, etc. The difference is that a job posting results in a very controlled number of applications, unlike movies where thousands of users can be provided a recommendation.

RecSys have held competitions to garner the attention of researchers in this domain and work closely with partners from the industry for solving real-world recommendation challenges. Challenges around the process of job recommendation were hosted during the years 2016 and 2017 (Abel,et al., 2017; Abel,et al., 2016). In 2016, Zibriczky et al.(2016) used a composition of 11 predictor instances as a solution to the challenge. He showed that based on forward predictor selection, item-neighbor methods and interaction data have great potential in improving offline accuracy. In 2017, Volkovs et al.,(2017) used a combination of content and neighbor-based models in their approach. They used user, item and user-item interaction features in Deep Neural Networks (DNN) and Gradient Boosting Machines (GBM), predicting in the output whether a user will positively interact with a job. They observed that due to input sparsity and feature ranges training DNNs were slow. DNNs were also sensitive to the choice of normalization when dealing with sparsity. GBMs worked well without any input pre-processing or normalization. Volkovs et al.,(2017) also solved the cold-start problem of missing user and item data through their approach. Liu et al.,(2016) showed the use of temporal learning and sequence modeling which captured complexities of user-item interactions to improve job recommendations.

Convolutional Neural Networks (CNN) and Recurrent Neural Networks (RNN) have also been used to solve the job recommendation problem. Zhu et al.,(2018) proposed PJFNN, a novel data driven, end-to-end model based on CNN. They used a Person-Job Fit Neural Network to learn the joint representations of Person-Job fitness from historical applications. Qin et al.,(2018) used an Ability-aware Person-Job Fit Neural Network, an ability aware model which combines RNN and four hierarchical ability aware attention strategies to determine Person-Job fit.

In our approach, we use sequence modeling with attention to capture nuances in the progression of job selection by the candidates. We pay special attention to skills as found in (Bastian ,et al., 2014) and propose novel ways to construct two representations of candidate and job skills. We also compose our approach using blended sub-recommendations that makes the final recommendation serendipitous and overcomes problems of cold-start in recommendations.

## 3 Experiments

Our approach uses a blend of machine learning models and other sub-recommendations to suggest jobs to candidates. Using machine learning models, we attempt to capture candidate's progression of job selection. Consequently, using machine learning models might also help to capture any latent motivations of the candidates while they have interacted with jobs. Recommendations from a machine learning model produce jobs that the candidate is most likely to click or interact with. The dataset to train the model was constructed using implicit and explicit feedback present in candidate-job interactions from our database. Explicit feedback is when the candidate clicks on a job to further expand its contents or clicks on the *apply* button to apply to a job. Implicit feedback is when a recruiter tags a candidate to a job. We can use features from candidate and job data as the input to any machine learning model and train it to predict 1 if the candidate will interact or 0 if the candidate won't interact with the job. We used several machine learning approaches as shown in *Section 3.3 Models*.

While machine learning methods attempt to capture the overall trends in data and the progression of job selection by the candidates, we found that job recommendations made only using machine learning methods are somewhat monotonous. For example, while it is common to recommend jobs requiring *Programming skills* to a *Software Developer*, this results in showing too many similar jobs. We experimented with several strategies to break the monotonicity. To capture the sentiment behind what makes a job exciting to a candidate, we needed to draw inspiration from real life scenarios. First, a candidate might ask for job recommendations from their peer group. Second, when a candidate applies for a job, it probably captures the specific interest of the candidate toward the job and similar jobs as opposed to the recruiter selecting the candidate for the job. We performed several experiments to capture the essence of these real life scenarios in our methodology and found that adding a small percentage jobs from a) jobs applied to by similar candidates and b) similar jobs applied to by the candidate in question could potentially make the recommendations serendipitous and might motivate candidates towards choosing jobs that excite them.

The techniques used in our blended approach also naturally solve the job and candidate cold-start problem. Due to the absence of progression, a new candidate may not aptly leverage the machine learning model for job recommendations, and neither can a new job be recommended by the model to any candidate. However, a new candidate can be shown jobs applied by similar candidates where job interaction data is present. Similarly, when comparing two jobs, a new job can get suggested when creating a recommendation based on similar jobs applied to by the candidate. Finally, we observed that using our blended approach increased the click-through rate (CTR) as a consequence of candidates interacting more frequently with the recommended jobs on our job web portal.

### 3.1 Methodology

The recommender system we developed is demonstrated in Figure 1. A candidate logs in to the *Job Web Portal* and their meta-data is forwarded to the *Recommendation Composer Module*. The *Recommendation Composer Module* then uses the features of the candidate to build a job filter using relaxed parameter values to extract a subset of relevant jobs. Consider a hypothetical example, if the professional experience of the candidate is 4 years, then the filter specifies minimum experience as 3 years and maximum experience as 5 years.

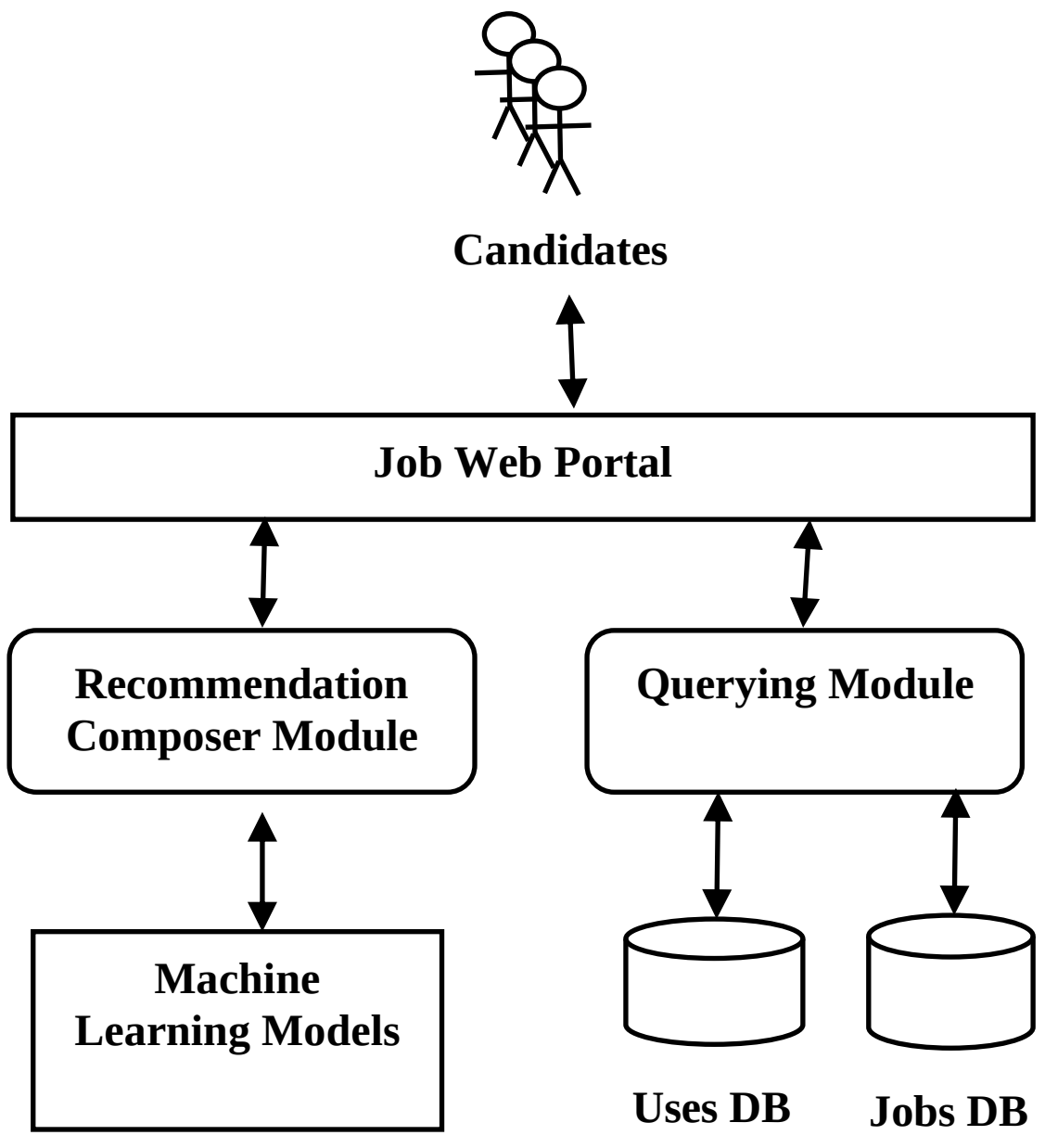

Figure 1: Proposed job recommendation system

The *Recommendation Composer Module* sends the job filter to the *Querying Module*. The *Querying Module* then presents the *Recommendation Composer Module* with the results obtained from the databases using the job filter. The role of the *Querying Module* is to construct queries according to the filters provided to it, fetch the relevant records from the databases and present the output in both raw and vectorized formats. The vectorized format can be directly used as input to machine learning models or for other vectorized computations. The raw format can be used to compose human readable recommendations. The *Recommendation Composer Module* generates sub-recommendations that are generated by different methods. Finally, the *Recommendation Composer Module* composes the final recommendation of jobs for the candidate's viewing. In order to learn the progression of job selection by candidates, we train a Bi-LSTM with attention model. The final recommendation is composed using a blended approach defined by the steps below.

**Step 1** *Creating a job filter:* Create a job filter using relaxed values in candidate features. This filter is submitted by the *Recommendation Composer Module* to the *Querying Module* that responds with a set of jobs, $J_{filtered}$. All sub-recommendation methods described in the next steps will use this reduced set of jobs, $J_{filtered}$, for computational efficiency.

**Step 2** *Checking interaction data:* Check if the candidate has an interaction history. If interaction history is present, then go to Step 3. Else go to Step 5.

**Step 3** *Applying machine learning model:* Fetch the job interaction history of the candidate. Using $J_{filtered}$ and the interaction history, the vectorized candidate and job features are used to predict the recommended jobs using a Bi-LSTM with attention model. An initial ranked recommendation is created using the decreasing order of *created-on* attribute of the job, $R_{machine\ learning}$. Figure 2 shows the architecture of the recommendations composed using Bi-LSTM with attention.

**Step 4** *Creating recommendations using non-machine learning methods - Similar Jobs:* Using $J_{reduced}$, find the set of jobs previously applied to by the candidate and select similar jobs where the cosine similarity score with other jobs is >= 0.70. Sort these jobs on the decreasing order of their *created-on* attribute and prepare a job recommendation list $R_{non\text{-}machine\ learning\ I}$. We can see here that this step assists in solving the job cold-start problem since a new job will be picked up if it is similar to the job being compared to.

**Step 5** *Creating recommendations using non-machine learning methods – Similar Candidates:* Using the candidate vector, select similar candidates where the cosine similarity score with other candidates is >= 0.80. From $J_{reduced}$, fetch the jobs applied by the similar candidates, sort them on the decreasing order of their *created-on* attribute and prepare a job recommendation list $R_{non\text{-}machine\ learning\ II}$. We can see here that this step aids in solving the candidate cold-start problem since interaction history of the candidate is not required.

**Step 6** *Blending Recommendations:* There are two ways to compose the final recommendation in this step. a) If $R_{machine\ learning}$ is non-empty, add all the jobs in $R_{machine\ learning}$ to the final recommendation, $R_{final}$. Next, choose 2 jobs from $R_{non\text{-}machine\ learning\ I}$ and $R_{non\text{-}machine\ learning\ II}$ respectively and insert them at random positions in $R_{machine\ learning}$ for every 10 jobs. b) If $R_{machine\ learning}$ is empty, alternately add jobs from $R_{non\text{-}}$ $_{machine\ learning\ I}$ and $R_{non\text{-}machine\ learning\ II}$ to $R_{final}$. It is obvious that jobs that the candidate has already applied to will not be included in $R_{final}$. We have jumbled the recommendations, thereby attempting to break the monotonicity of machine learning recommendations.

**Step 7** *Accounting for edge cases:* This step accounts for the edge case where the final recommendation, $R_{final}$, is empty. The probable causes could be an independent or combined effect of a) new candidates or jobs added to the system that are completely new and are distant from the threshold values we have assumed in the respective cases b) the candidate has already applied to all the recommended jobs. In this case, we compose the recommendation using overlap between the candidate and jobs using $J_{reduced}$. We use cosine similarities between the skills of the candidate and those stated by the jobs and perform some fuzzy matching of other candidate-job features like overlap of *experience*, *industry* and *job-title.* Also, a scheduled task periodically keeps a count if a job appeared in $J_{reduced}$ and was still not shown to the candidate. When this count exceeds the threshold (50) it inserts the respective jobs into random positions in the final recommendation thereby preventing some cases when a job could never get recommended.

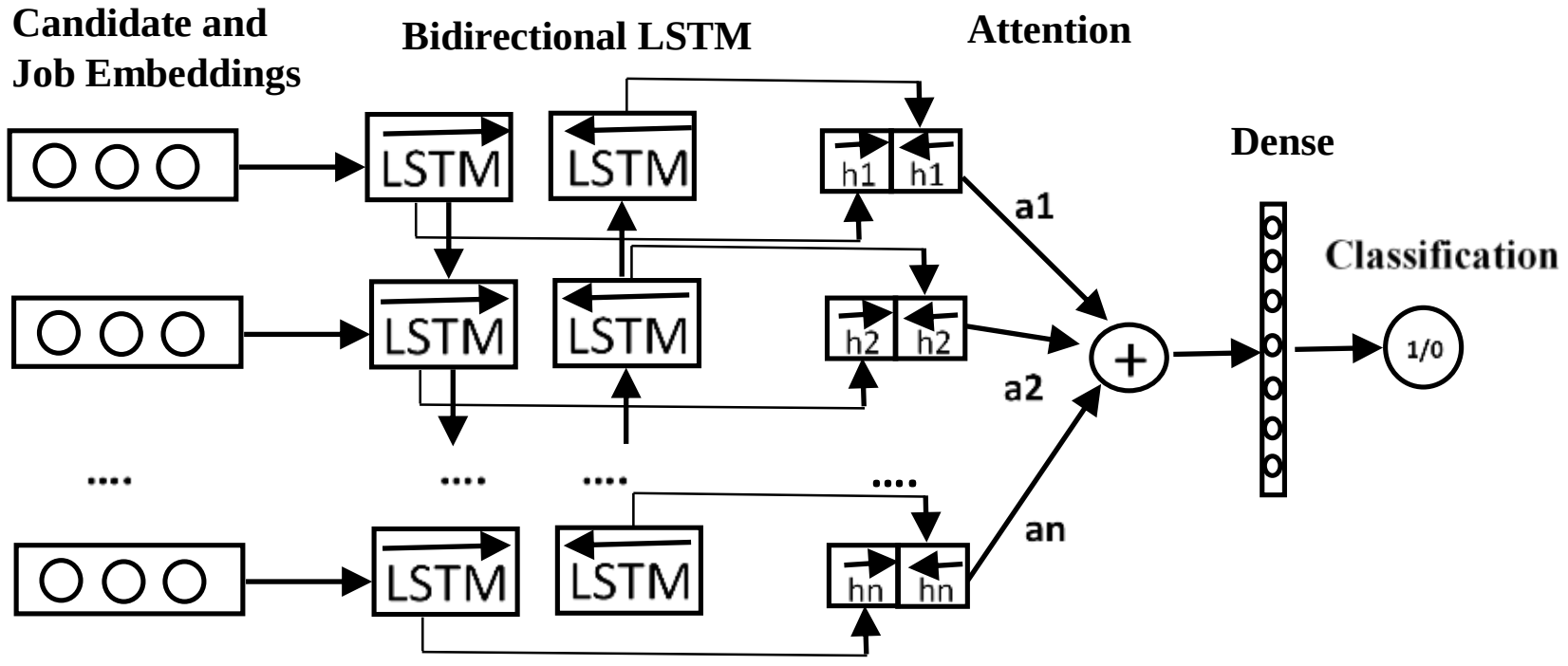

Figure 2: Bi-LSTM with Attention architecture for job recommendation

## 3.2 Dataset and Feature Selection

We construct the dataset for our experiments using data from our organization's database. The dataset contains 4208 distinct candidates and 2334 distinct jobs. The latest date of job that any candidate has applied for is from March 2019 and the earliest date of the job that any candidate has applied for is from April 2014. We select only those candidates who have interacted within this time span. The total interactions between the candidates and the jobs are 1125776. Interactions represent a) recruiter tagging a candidate for a job, b) candidate clicking on a job to further expand its contents and c) candidate clicking the *apply* button to start their job application process. These are all *favorable* or *positive* outcomes and we assume that collectively, the candidate has *clicked* on these jobs. While searching for a job a candidate may be shown jobs which the candidate may choose to ignore. These form the *negative* outcomes. For our machine learning models, this translates into a classification problem where we try to predict a positive (1) outcome or a negative (0) outcome generated by a user for any given job. The dataset and interaction data have been summarized in Table 1 and Table 2 respectively.

| | |
|---|---|
| Distinct Candidates | 4208 |
| Distinct Jobs | 2334 |
| Positive Interactions | 316498 |
| Negative Interactions | 809278 |

Table 1: Dataset summary

| **Positive Interactions** | |
|---|---|
| Recruiter tagged a candidate to a job | 215218 |
| Candidate expanded a Job | 72794 |
| Candidate applied to a Job | 28486 |
| Total | 316498 |
| **Negative Interactions** | |
| Candidate ignored job shown | 809278 |

Table 2: Interaction Data Summary

We shortlisted 9 features from each candidate, 11 features from each job and 1 common feature that adds up to a total of 21 features. We split the data into 70%, 20% and 10% for training, testing and validating sets respectively. To represent candidate and job skills in our dataset, the word embeddings learned by the Word2Vec model is used. The dimensionality of the word vectors is 20, the training algorithm is continuous Bag-of-Words, the window size is 5 and min_count is 5. A T-SNE plot of the final Word2Vec model with some sample skills is shown in Figure 3.

We observe that while skills are an important denominator for matchmaking, sometimes semantic information from skills alone might not suffice for ideal matchmaking. This is because there are several ways in which candidates and recruiters define skills and competencies. Sometimes one skill may portray a collective meaning for several constituent skills. For instance, a candidate who mentions *Full Stack Developer* as a skill might have latent competencies in *Microservices*, *Web Development*, *Javascript*, *Angular*, etc. Similarly, a recruiter posting a job having the skill requirements of a *Web Developer* may also be interested in candidates having competencies in *HTML*, *Microservices*, *Javascript* and so on. We assumed that using *Latent Competency Group Similarity* (defined in the subsequent paragraph) between a job and a candidate along with skills would assist our machine learning models to make better inferences.

Competency groups are domain specific aggregation of skills. For example, skills such as *linear regression*, *natural language processing*, *deep learning*, *data visualization* and so on belong to the *machine learning* competency group. *Data* visualization can also belong to the competency group *data science*, hence a skill can appear in multiple competency groups. A recruiter can just state *machine* learning as a required skill for a job and a deserving candidate could express their skills using one or more keywords. We attempt to “reveal” the overlap of domains between jobs and candidates using competency groups and hence named this as Latent Competency Groups. We gathered a team of data analysts and subject matter experts to create the latent competency groups. Everyone involved was compensated for the task. The final reviewed latent competencies included 100 groups.

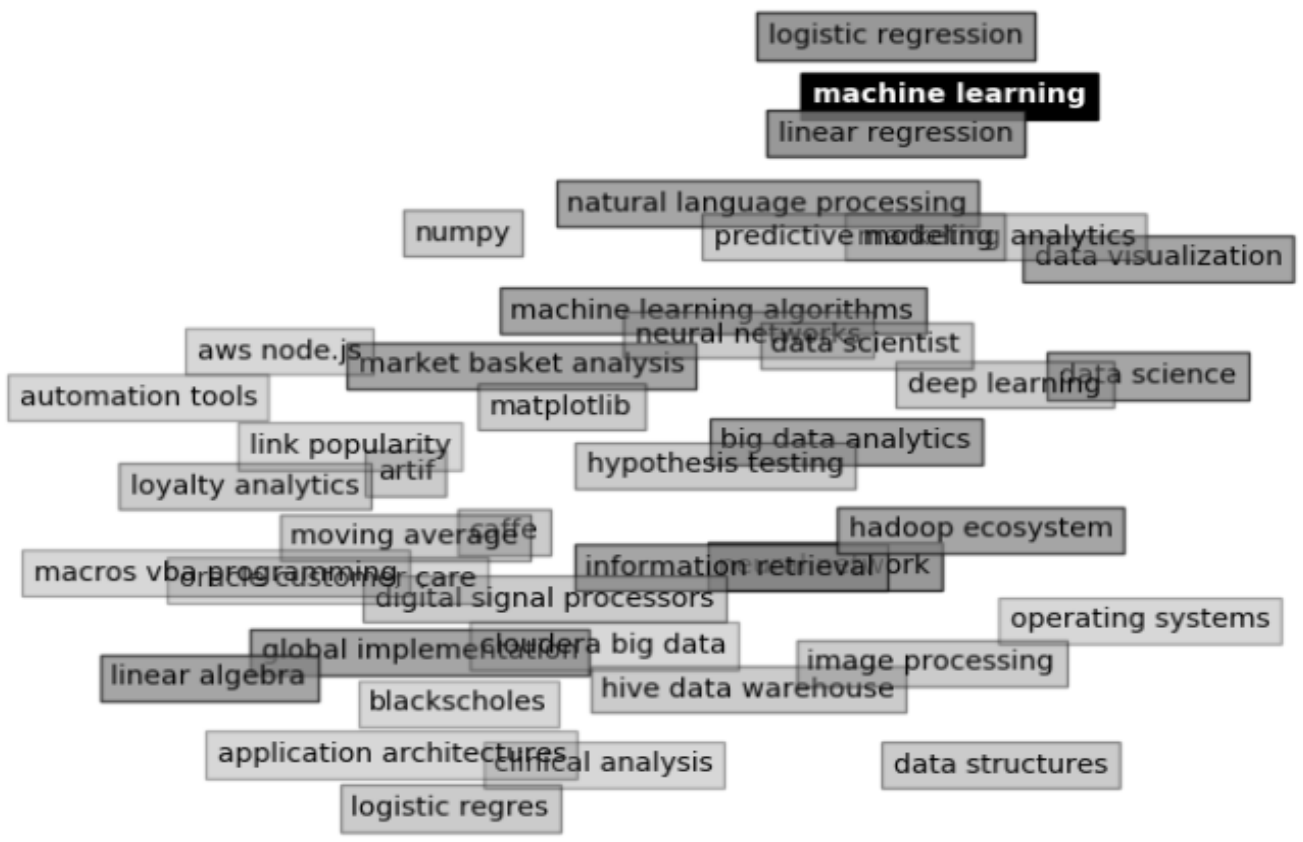

Figure 3: T-SNE Plot of Word2Vec on Sample Skills

We represent the skills of a candidate or a job by a vector where each dimension represents a latent competency group. For each candidate or job, first, a vector $\boldsymbol{V}$ of size 100 is created and initialized with 0’s. Each index in this vector represents a group. For each skill, the associated groups are identified, and 1 is added to the corresponding indices in $\boldsymbol{V}$. Then, the values in $\boldsymbol{V}$ are normalized between 0 and 1. Next, *Latent Competency Group Similarity* between a candidate and a job is computed which is the cosine similarity value of $\boldsymbol{V_c}$ and $\boldsymbol{V_j}$, where $\boldsymbol{V_c}$ represents the candidate latent competency group vector and $\boldsymbol{V_j}$ represents the job latent competency group vector.

The expansion of skills into latent competency groups using the above methodology attempts to capture latent skills that humans can infer but may remain hidden for machine learning models due to the brevity used by recruiters and candidates while mentioning skills.

## 3.3 Models

Our methodology composes job recommendations using several strategies where a sub-task is to choose the best machine learning model that captures the progression of job selection by candidates. We approach this sub-task as a classification problem where 1 represents a candidate interacted with a job or a recruiter tagged a candidate to a job and 0 represents jobs that were shown but not interacted with by the candidate in our job web portal.

We experimented with several machine learning algorithms that included both tree-based approaches and deep neural networks.

| Machine Learning Models | Hyperparameters |
|---|---|
| Random Forest | Criterion: gini<br>N_estimators: 300 |
| XGBoost | N_estimators: 500 |
| Bi-Directional LSTM with Attention | Timesteps: 2<br>Hidden Layers: 2 (Nodes: 128, 64)<br>Optimizer: Adam<br>Dropout: 20% |

Table 3: Machine Learning Model Hyperparameters

We chose Random Forests and XGBoost that are tree-based approaches and these methods performed well. However, Bi-LSTM with attention gave us more accurate results. We used these algorithms from the *scikit-learn* Python module. We used grid search with cross-validation for choosing the best hyperparameters. The hyperparameters we used for the different models are shown in Table 3.

### 3.4 Using Progression of Job Applications by Candidates through Bi-LSTM

The aim of this experiment was to imitate a recruiter who assimilates a candidate's progression for determining relevant jobs for them. For instance, as a candidate progresses in their career by changing jobs, growing into a higher role, moving to a different place, they update their resume accordingly. A recruiter, perhaps, screens a candidate by trying to understand the major advances in their career and how the candidate would make a good fit in the current role.

When a candidate interacts with different jobs over time, some of their latent preferences are hidden in these interactions. The training data is modeled to capture these changing job-preferences of candidates over time, not to mention any explicit changes in candidate's attributes over time, like skills and location, are also captured (see Equation 1). In Equation 1, during training the Bi-LSTM, $CJ_1$ is a candidate-job pair where candidate C positively interacts with job $J_1$. $C'J_2$ represents the same candidate (with updated attributes) at a later point in time and job $J_2$. The target variable is 1 or 0, depending on whether C' positively or negatively interacts with $J_2$ respectively.

$$CJ_1 \longrightarrow C'J_2 \longrightarrow 1 / 0 \qquad (1)$$

Where, 1 – C' Interacts with $J_2$, 0 – C' doesn't Interacts with $J_2$ and Timesteps = 2. During testing, while predicting whether a candidate will positively interact with a job, attributes of the last job applied to by the candidate ($J_1$) and the attributes of the candidate's profile (C), at the time C positively interacted with $J_1$, forms the first timestep. The job in question ($J_2$) and the candidate's current profile (C') are used to form the second timestep. The model predicts if C' will positively or negatively interact with $J_2$.

We chose Bi-LSTM since its architecture can leverage both past as well as future candidate-job interactions (progressions) to learn some of the latent job preferences of candidates and predict if they will likely interact with given jobs.

## 4 Results

We used several machine learning algorithms to learn the progression of job selection by candidates, and the results have been summarized in Table 4. The Bi-LSTM with attention model gave us the best results. The diagram showing the components of the Bi-LSTM model is shown in . Bi-LSTM provided superior results due to its ability to learn progression in the form of sequences and use interaction information from the past to predict future outcomes.

We also computed feature importance using the Random Forest model. Skills and OrganizationID were the most predictive candidate features. Skills and Industry Name were the most predictive job features. Latent Competency Group Similarity was also highly predictive.

| Model | Accuracy | Precision | | Recall | | F1-Score | |
|---|---|---|---|---|---|---|---|
| | | **Class 0** | **Class 1** | **Class 0** | **Class 1** | **Class 0** | **Class 1** |
| Random Forest | 91.49 | 93.96 | 80.90 | 95.95 | 72.58 | 94.80 | 76.51 |
| XGBoost | 91.43 | 94.17 | 79.04 | 95.30 | 75.03 | 94.73 | 76.99 |
| ANN | 91.53 | 93.75 | 81.09 | 96.13 | 72.56 | 94.93 | 76.58 |
| Bi-LSTM with Attention | 92.02 | 95.93 | 82.42 | 97.52 | 75.13 | 95.72 | 78.61 |
| Encoder Decoder | 71.63 | 87.88 | 34.88 | 75.32 | 55.99 | 42.98 | 42.98 |

Table 4: Results

We found that when there were too many jobs to recommend, all of which had similar criteria, the recommendations became monotonous. It motivated us to dive deeper into the job application process of the candidates and take inspiration from real life scenarios and attempt to make our job recommendations serendipitous for the candidate. We also needed to address the job and candidate cold-start problems. Hence, we introduced a blended approach where we used non-machine learning based techniques. We added a) jobs applied to by similar candidates and b) similar jobs applied to by the candidate, in small proportions to the recommendations from the Bi-LSTM with attention model. The complete process of constructing the blended recommendation along with the choice of similarity comparison method and threshold values have been described in Section *3.1 Methodology*.

We found significant improvement in our job web portal with the blended approach and saw a relative increase of 63% in click-through rates (CTR). The results are statistically significant by chi-square test at $p < .01$ and the candidate-job datasets (pre-model and post-model) belong to the same population for key attributes such as experience, domain, industry, and function at $p < .01$.

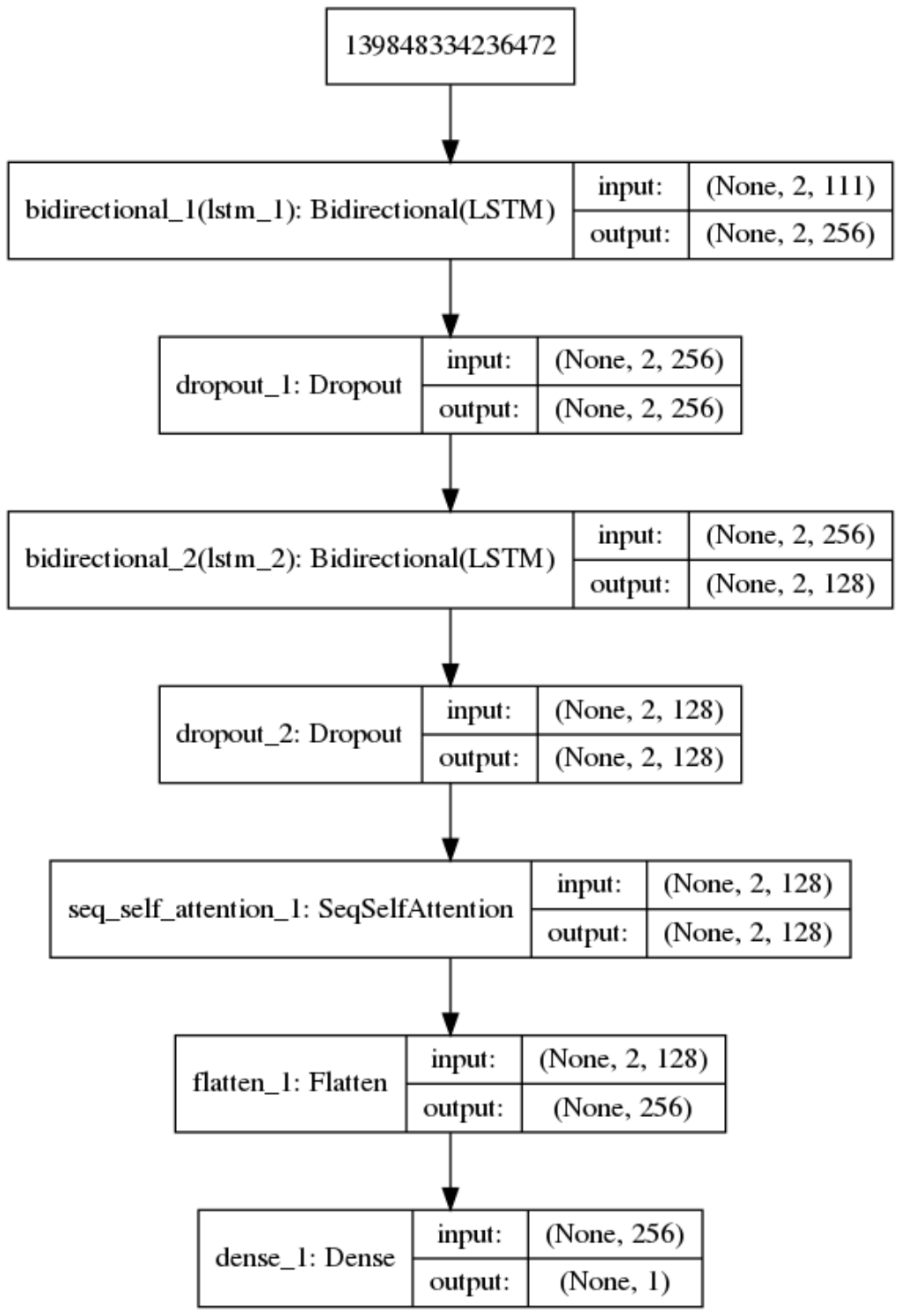

Figure 4: Bi-LSTM model with Attention

## 5 Conclusion and Future Work

This paper demonstrates a novel blended approach that leverages progression of job selection by candidates and attempts to make job recommendations serendipitous. Using blended methods, recommendations suggested to candidates are based on their interaction history with jobs, along with jobs that are a) similar to the other jobs applied by the candidate and b) applied by similar candidates. Our approach naturally solves the candidate and job cold-start problem in the absence of interaction data. We also demonstrated the use of latent competency groups which expand the job skill requirements and the candidate skills thereby attempting to reveal latent competencies and achieve more coverage in the skill domain. Using our methodology, we see a relative increase in click-through rates of candidates visiting our portal and applying for jobs.

As part of the future work, we plan to use features of similar candidates and jobs in sequence information. As of now, recommendation using similar candidates and jobs forms part of non-machine learning based recommendations and the initial results seem promising. Finally, it would be interesting to extend our methodology to other recommender systems.

## References

Mathieu Bastian, Matthew Hayes Hayes, William Vaughan, Sal Uryasev and Christopher Lloyd. 2014. LinkedIn skills: large-scale topic extraction and inference. In Proceedings of the 8th ACM Conference on Recommender systems (pp. 1-8). ACM.

Paul Covington, Jay Adams and Emre Sargin. 2016. Deep neural networks for youtube recommendations. In Proceedings of the 10th ACM conference on recommender systems (pp. 191-198). ACM.

Carlos A. Gomez-Uribe and Neil Hunt. 2016. The netflix recommender system: Algorithms, business value, and innovation. ACM Transactions on Management Information Systems (TMIS), 6(4), 13.

Shumpei Okura, Yukihiro Tagami, Shingo Ono and Akira Tajima. 2017. Embedding-based news recommendation for millions of users. In Proceedings of the 23rd ACM SIGKDD International Conference on Knowledge Discovery and Data Mining (pp. 1933-1942). ACM.

Ahmed Elsafty, Martin Riedl and Chris Biemann. 2018.Document-based Recommender System for Job Postings using Dense Representations. In Proceedings of the 2018 Conference of the North American Chapter of the Association for Computational Linguistics: Human Language Technologies, Volume 3 (Industry Papers) (pp. 216-224).

Alexander Tuzhilin and Gediminas Adomavicius. 2005. Toward the next generation of recommender systems: A survey of the state-of-the-art and possible extensions. IEEE Transactions on Knowledge & Data Engineering, (6), 734-749.

Fabian Abel, Yashar Deldjoo, Mehdi Elahi and Daniel Kohlsdorf. 2017. Recsys challenge 2017: Offline and online evaluation. In Proceedings of the Eleventh ACM Conference on Recommender Systems (pp. 372-373). ACM.

Dawei Chen, Cheng Soon Ong and Aditya Krishna Menon. 2019. Cold-start playlist recommendation with multitask learning. arXiv preprint arXiv:1901.06125.

Andrew I. Schein, Alexandrin Popescul, Lyle H. Ungar and David M. Pennock. 2002. Methods and metrics for cold-start recommendations. In Proceedings of the 25th annual international ACM SIGIR conference on Research and development in information retrieval (pp. 253-260). ACM.

M Jiang, Y Fang, H Xie, J Chong and M Meng. 2019. User click prediction for personalized job recommendation. World Wide Web, 22(1), 325-345.

K Kenthapadi, B Le and G Venkataraman. 2017. Personalized job recommendation system at linkedin: Practical challenges and lessons learned. In Proceedings of the Eleventh ACM Conference on Recommender Systems(pp. 346-347). ACM.

Fabien Abel, A Benczúr, D Kohlsdorf, M Larson and Róbert Pálovics. 2016. Recsys challenge 2016: Job recommendations. In Proceedings of the 10th ACM Conference on Recommender Systems (pp. 425-426). ACM.

Tomas Mikolov, Ilya Sutskever, Kai Chen Greg S. Corrado and Jeff Dean. 2013. Distributed representations of words and phrases and their compositionality. In Advances in neural information processing systems (pp. 3111-3119).

Quoc Le and Tomas Mikolov. 2014. Distributed representations of sentences and documents. In International conference on machine learning (pp. 1188-1196).

Dávid Zibriczky. 2016. A combination of simple models by forward predictor selection for job recommendation. In Proceedings of the Recommender Systems Challenge (p. 9). ACM.

Maksims Volkovs, Guang Wei Yu and Tomi Poutanen. 2017. Content-based neighbor models for cold start in recommender systems. In Proceedings of the Recommender Systems Challenge 2017 (p. 7). ACM.

Kuan Liu, Xing Shi, Anoop Kumar, Linhong Zhu and Prem Natarajan. 2016. Temporal learning and sequence modeling for a job recommender system. In Proceedings of the Recommender Systems

Challenge (p. 7). ACM.

Chen Zhu, Hengshu Zhu, Hui Xiong, Chao Ma, Fang Xie, Pengliang Ding and Pan Li. 2018. Person-Job Fit: Adapting the Right Talent for the Right Job with Joint Representation Learning. ACM Transactions on Management Information Systems (TMIS), 9(3), 12.

Chuan Qin, Hengshu Zhu, Tong Xu, Chen Zhu, Liang Jiang, Enhong Chen and Hui Xiong. 2018. Enhancing person-job fit for talent recruitment: An ability-aware neural network approach. In The 41st International ACM SIGIR Conference on Research & Development in Information Retrieval (pp. 25-34). ACM.